\documentclass[twocolumn,aps]{revtex4-1}
\usepackage{bm}
\usepackage{epsfig}
\usepackage{amsmath}

\def\lazz{\mathrel{\mathchoice {\vcenter{\offinterlineskip\halign{\hfil
$\displaystyle##$\hfil\cr<\cr\sim\cr}}}
{\vcenter{\offinterlineskip\halign{\hfil$\textstyle##$\hfil\cr<\cr\sim\cr}}}
{\vcenter{\offinterlineskip\halign{
\hfil$\scriptstyle##$\hfil\cr<\cr\sim\cr}}}
{\vcenter{\offinterlineskip\halign{\hfil$\scriptscriptstyle##
$\hfil\cr<\cr\sim\cr}}}}}

\def\be{\begin{equation}}
\def\lan{\left\langle}
\def\ran{\right\rangle}
\def\ee{\end{equation}}
\def\barr{\begin{array}}
\def\earr{\end{array}}

\def\nn8{\\}
\def\l{\left}
\def\r{\right}
\def\dis{\displaystyle}
\def\ed{\end{document}}

\def\cs{{\bf s}}

\begin{document}

\title{Probability Distribution of the Ratio of Consecutive Level Spacings\\ in
Interacting Particle Systems}

 \author{N. D. Chavda$^a$ and V. K. B. Kota$^b$}
 \affiliation{$^a$Applied Physics Department, Faculty of Technology and
Engineering,\\ M. S. University of Baroda, Vadodara 390 001, India\\
$^b$Physical Research Laboratory, Ahmedabad 380 009, India}

\begin{abstract}

We study the probability distribution of the  ratio of consecutive level
spacings for embedded one plus two-body random matrix ensembles with and without
spin degree of freedom and for both fermion and boson systems. The agreement
between the numerical results and the recently derived analytic form for the
distribution and other related quantities is found to be close. This
establishes conclusively that local level fluctuations generated by embedded
ensembles follow the results of classical Gaussian ensembles.

\end{abstract}

\date{\today}
\maketitle


\section{Introduction}
\label{sec:1}
Large number of investigations in the last two decades have established that
Embedded Gaussian orthogonal ensembles of one plus two-body interactions
[EGOE(1+2)] operating in many particle spaces \cite{vkbk01,Gomez,Weiden}, apply in a
generic way to isolated finite interacting many-particle quantum systems such as
nuclei, atoms, quantum dots, small metallic grains, interacting spin systems
modeling  quantum computing core and so on.  For sufficiently strong
interaction, EGOEs exhibit  average-fluctuation separation in eigenvalues with
the smoothed eigenvalue  density being a corrected Gaussian and the local
fluctuations are of GOE type \cite{Gomez,Br-81,Lec-08}.

The classical Gaussian Orthogonal Ensemble (GOE), Gaussian Unitary Ensemble (GUE)
and Gaussian Symplectic ensemble (GSE) are ensembles of multi-body ($m$-body for
$m$-particle systems) interactions. Therefore they are unrealistic for
understanding quantum many-body chaos as interactions for most finite quantum
systems are largely one- plus two-body in  character (one-body part corresponds to
a mean-field). The concept of Embedded Ensembles (EE) precisely takes care of the
one- plus two-body nature of the  interactions. In this paper all the discussion
is restricted to GOE version of EE. For $m$ spin-less fermions
occupying say $N$ single particle (sp) states, with the Hamiltonian ($H$) matrix in
two-particle spaces represented by GOE and  then constructing the many-particle
$H$ matrix, with the $m$-particle basis states being direct products of sp states,
 gives the Embedded Gaussian Orthogonal Ensemble of
two-body interactions [EGOE(2)] in $m$-particle spaces. Similarly, for
interacting spin-less boson systems, the Embedded Gaussian Orthogonal Ensemble of
two-body interactions can be defined and to distinguish these from those of
fermion systems, they are denoted by BEGOE(2) \cite{asaga}. Addition of the
mean-field, one-body part, gives EGOE(1+2) and BEGOE(1+2) for fermion and boson
systems respectively.  EGOE(1+2) can be defined for fermions or bosons with spin
degree of freedom and also with many other symmetries \cite{Gomez,Kota}. It is
useful to note that EGOE(1+2)s [BEGOE(1+2)s] reduce to EGOE(2)s [BEGOE(2)s] in
the strong interaction limit.

Normally one uses the Nearest Neighbor Spacing Distribution (NNSD) $P(S)dS$
for establishing GOE nature of level fluctuations. It is well known that if the
system is in integrable domain, corresponding to the regular behavior of the
system, the form of the NNSD is close to the Poisson distribution [$P(S) = \exp (
-S)$]. However, if the system is chaotic, NNSD is described by the Wigner surmise
which is essentially the GOE result [$P(S) = (\pi/2) S \exp (-\pi S^2/4)$]. In
constructing NNSD for a given set of energy levels (or eigenvalues),  the
spectra have to be unfolded to remove the variation in the density of
eigenvalues \cite{Br-81,Haake}. For the spectra generated by a random matrix
ensemble, the unfolding of the spectrum can be done in two different ways. One
of them being  spectral unfolding, i.e. spectrum of each individual member of
the ensemble is unfolded separately and then ensemble averaged NNSD is
constructed. A second method is  ensemble unfolding and here a single unfolding
function is used for all the members. For EGOE, it is well known that spectral
unfolding gives GOE fluctuations while there are deviations from GOE when
ensemble unfolding is used \cite{Br-81,BF71}. Because of this, in the past there
was serious confusion with regard to the usefulness of EE; see the discussion on
page 424 of \cite{Br-81} and also \cite{Bo-74}.  Another minor issue is,
although eigenvalue density for EGOEs is close to Gaussian form, many groups use
polynomial of a high degree for unfolding purposes; see for example
\cite{Retamosa}. Therefore, to conclusively establish that EGOEs generate GOE
fluctuations, it is important to use fluctuation measures that are independent
of the unfolding function and unfolding procedure. The purpose of the present
letter is to report the results of the analysis of level fluctuations in EGOEs
using a measure introduced  recently that is independent of unfolding the
spectrum.

Oganesyan and Huse \cite{Huse2007} in 2007 considered the distribution of the ratio
of consecutive level spacings of the energy levels which requires no unfolding
as it is independent of the form of the density of the energy levels. This
distribution allows a more transparent comparison with experimental results than
the traditional level spacing distribution. This measure was used to quantify
the distance from integrability on finite size lattices
\cite{koll2010,Coll2012}, and also to investigate numerically many-body
localization \cite{Huse2007,OPH2009,Pal-10,Iyer-12}.  More importantly, recently
Atas et. al \cite{ABGR-2013}, derived expressions for the probability
distribution of the ratio of two consecutive level spacings for the classical
GOE, GUE and GSE ensembles of random matrices. These  expressions, called
Wigner-like surmises, are shown to be very accurate when compared to numerical
calculations in the large matrix size limit. Also, results from a quantum
many-body lattice model and from zeros of the Riemann zeta function  are shown
to be in excellent agreement with the analytical formulas derived.  Going beyond
these, in the present letter we show that  the probability distribution for
ratio of two consecutive level spacings and also the related averages for many
different EGOEs are very close to the GOE formulas. This then confirms
conclusively that EGOE level fluctuations follow GOE. Now, we will give a
preview.

Analytical results for GOE for the probability distribution of the ratio of
consecutive level spacings and related averages are briefly discussed in
Section \ref{sec:2}. The five different EGOEs used in the present analysis are described
in Section \ref{sec:3}. The numerical EGOE (and BEGOE) results of the probability
distribution for ratio of consecutive spacings and related averages are presented in Section
\ref{sec:4}. Finally, Section \ref{sec:5} gives concluding remarks.

\section{Probability Distribution of the Ratio of Consecutive Level Spacings}
\label{sec:2}
Let us consider an ordered set of eigenvalues (energy levels) $E_n$, where
$n=1,2,...,d$. The Nearest-Neighbor Spacings is given by $s_n = E_{n+1} -
E_{n}$. Then, the ratio of two consecutive level spacings is $r_n=s_{n+1}/s_n$.
The probability distribution for consecutive level spacings is denoted by
$P(r)dr$. If the system is in integrable domain, then NNSD is Poisson and $P(r)$
is [denoted by $P_P(r)$],
\be
P_P(r)=\dis\frac{1}{(1+r)^2}\;.
\label{eq1}
\ee
Similarly, for GOE, derived using $3 \times 3$ real symmetric matrices,
the $P(r)$ is given by Wigner-like surmise \cite{ABGR-2013},
\be
P_W(r)= \frac{27}{8} \frac{r+r^2}{(1+r+r^2)^{5/2}} \;.
\label{prw}
\ee
It is also suggested in \cite{ABGR-2013} that the difference $\delta P(r)= P(r)
- P_W(r)$ between numerics and the surmise (\ref{prw}) can be approximated by
the following expression,
\be
\delta P(r) = \frac{C}{(1+r)^2} \left[ \l(r+\frac{1}{r}\r)^{-1} - 2\frac{\pi
-2}{4-\pi} \l(r+\frac{1}{r}\r)^{-2}  \right]\;,
\label{diff}
\ee
where the parameter $C$ is to be obtained by fitting the expression $P(r) =
P_W(r) + \delta P(r)$. In addition to $r_n$, Oganesyan and Huse \cite{Huse2007}
considered the distribution of the ratios $\tilde{r}_n$ defined by
\be
\tilde{r}_n=\frac{\min(s_n,s_{n-1})}{\max(s_n,s_{n-1})}=\min(r_n,1/r_n)
\label{eq.rn}
\ee
As pointed out in \cite{ABGR-2013}, it is possible to write down $P(\tilde{r})$
given $P(r)$. In practice, it is also useful to consider $\lan r\ran$, the
average value of $r$. For GOE we have $\lan r \ran= 1.75$ and for Poisson it is
$\infty$. However, $\lan \tilde{r}\ran=0.536$ for GOE and $0.386$ for Poisson.
We will use $P(r)$, $\lan r \ran$ and $\lan \tilde{r}\ran$ in the analysis of
energy levels presented in Section \ref{sec:4}.

\section{Embedded ensembles for fermion and boson systems with and without spin
degree of freedom}
\label{sec:3}

In the present study five different EGOEs are employed. For spin-less fermion and
boson systems with a mean-field and two-body interactions  we have EGOE(1+2) and
BEGOE(1+2) respectively \cite{vkbk01,cpk-2003} and the Hamiltonian  $H = h(1) +
\lambda \{V (2)\}$. Here, $h(1)$ is the mean-field one-body part defined by
single particle (sp) energies $\epsilon_i$ and $\{V (2)\}$ represents EGOE(2) or
BEGOE(2), i.e. $V(2)$ matrix in two-particle spaces is represented by GOE with
matrix elements variance unity. The parameter $\lambda$ is the strength of the
two-body interaction in units of average spacing $\Delta$  of the single sp
levels. Note that $\{---\}$ denotes ensemble with matrix elements variance unity
in two-particle spaces. Going beyond spin-less systems, we have considered three
embedded ensembles with spin degree of freedom. For fermions with spin $\cs=1/2$
degree of freedom, we have EGOE(1+2)-$\cs$ \cite{ksc-2006}. Here, the
interaction $V(2)$ will have two parts as the two particle spin $s = 0$ and $1$
giving $H=h(1)+\lambda_0 \{V^{s=0}(2)\}+\lambda_1 \{V^{s=1}(2)\}$. Similarly,
for two species boson systems it is possible to consider bosons with a
fictitious ($F$) spin $1/2$ degree of freedom. Then, we have BEGOE(1+2)-$F$
\cite{vyas-12}.  Again here also, the interaction will have two parts as the two
particle $F$-spin $f = 0, 1$.  Therefore for BEGOE(1+2)-$F$, $H =h(1)+\lambda_0
\{V^{f=0}(2)\}+\lambda_1 \{V^{f=1}(2)\}$. Note that here and also for
EGOE(1+2)-$\cs$, the sp levels will be doubly  degenerate. Finally, we have also
considered bosons spin one degree of freedom, i.e. BEGOE(1+2)-$S1$ \cite{Deota}.
Here, the interaction will have three parts as the two particle spin $s = 0$,
$1$ and $2$ and therefore $H =h(1)+\lambda_0 \{V^{s=0}(2)\}+\lambda_1
\{V^{s=1}(2)\} + \lambda_2 \{V^{s=2}(2)\}$ with sp levels defining $h(1)$
triply degenerate. In all the five ensembles, without loss of generality, we
choose the average spacing between the sp levels to be unity so that all
$\lambda$'s are unit-less.

The following choices are made for constructing the five EGOEs in many particle
spaces:

\begin{enumerate}

\item {EGOE(1+2) for $m=6$ fermions in $N=12$ sp states with $H$ matrix of
dimension 924. The sp energies are chosen as $\epsilon_i = i +
1/i$, $i = 1, 2, . . . , 12$ and the interaction strength $\lambda = 0.1$; see
Ref. \cite{vkbk01} for details.}

\item {EGOE(1+2)-$\cs$ for $m=6$ fermions occupying $\Omega=8$ sp levels (each
doubly degenerate) with total spin $S=0$ and $S=1$ giving the $H$ matrices of
dimensions 1176 and 1541 respectively. The sp energies are chosen as
$\epsilon_i = i + 1/i$, $i = 1, 2, . . . , 8$ and the interaction strength
$\lambda =\lambda_0 =\lambda_1 = 0.1$; see Ref. \cite{ksc-2006, mkc-2010}for
details.}

\item {BEGOE(1+2) for $m=10$ bosons in $N = 5$ sp states with $H$ matrix of
dimension 1001. The sp energies are chosen as $\epsilon_i = i + 1/i$, $i = 1,
2, . . . , 5$ and the interaction strength $\lambda = 0.03$; see
Ref.\cite{cpk-2003,ckp-2012} for details.}

\item {BEGOE(1+2)-$F$ for $m=10$ bosons occupying $\Omega = 4$ sp levels (each
doubly degenerate) with total $F$-spin $F=2$ and $F=F_{max}=5$ giving the $H$
matrices of dimensions 750 and 286.  The sp energies are chosen as
$\epsilon_i = i + 1/i$, $i = 1, 2, 3, 4$ and the interaction strength
$\lambda =\lambda_0 =\lambda_1= 0.05$; see Ref.\cite{vyas-12} for details.}

\item {BEGOE(1+2)-$S1$ for $m=8$ bosons occupying $\Omega = 4$ sp levels (each
triply degenerate) with total spin $S=4$ giving the $H$ matrix of dimension
1841. The sp energies are chosen as $\epsilon_i = i + 1/i$, $i
= 1, 2, . . . , 4$ and the interaction strength $\lambda = \lambda_0 =\lambda_1
= \lambda_2= 0.2$; see Ref.\cite{Deota} for details.}

\end{enumerate}

\noindent All the ensembles considered in the analysis have 500 members. Let us
add that the $\lambda$ values in the ensemble calculations are chosen such that
the system is in Gaussian domain, i.e. the state density and local density of
states will be close to Gaussian in form and level and strength fluctuations
(when an appropriate unfolding function is used) exhibit GOE fluctuations. Now
we will turn to the numerical results.

\section{Numerical results for examples from EGOE(1+2), BEGOE(1+2),
EGOE(1+2)-$\cs$, BEGOE(1+2)-$F$ and BEGOE(1+2)-$S1$}
\label{sec:4}

\begin{figure}
\begin{center}
\includegraphics[width=0.7\linewidth]{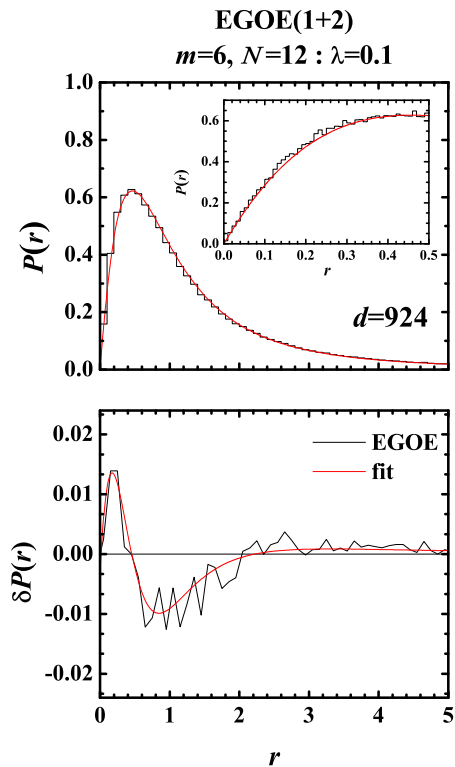}
\end{center}

\caption{Histogram, in the upper panel, represents probability distribution of
the  ratio of consecutive level spacings (represented by $P(r)$) for a 500
member EGOE(1+2) ensemble. The red smoothed curve is due to the surmise $P_W(r)$
given by Eq. (\ref{prw}). In the lower panel shown is difference $\delta
P(r)=P(r)-P_W(r)$ between ensemble average and the surmise $P_W(r)$. The red
smoothed curve is obtained by fitting Eq. (\ref{diff}). In the inset figure in the upper
panel  shown are results for $P(r)$ for $r \leq 0.5$.}

\label{egoe}
\end{figure}

\begin{figure}
\begin{center}
\includegraphics[width=\linewidth]{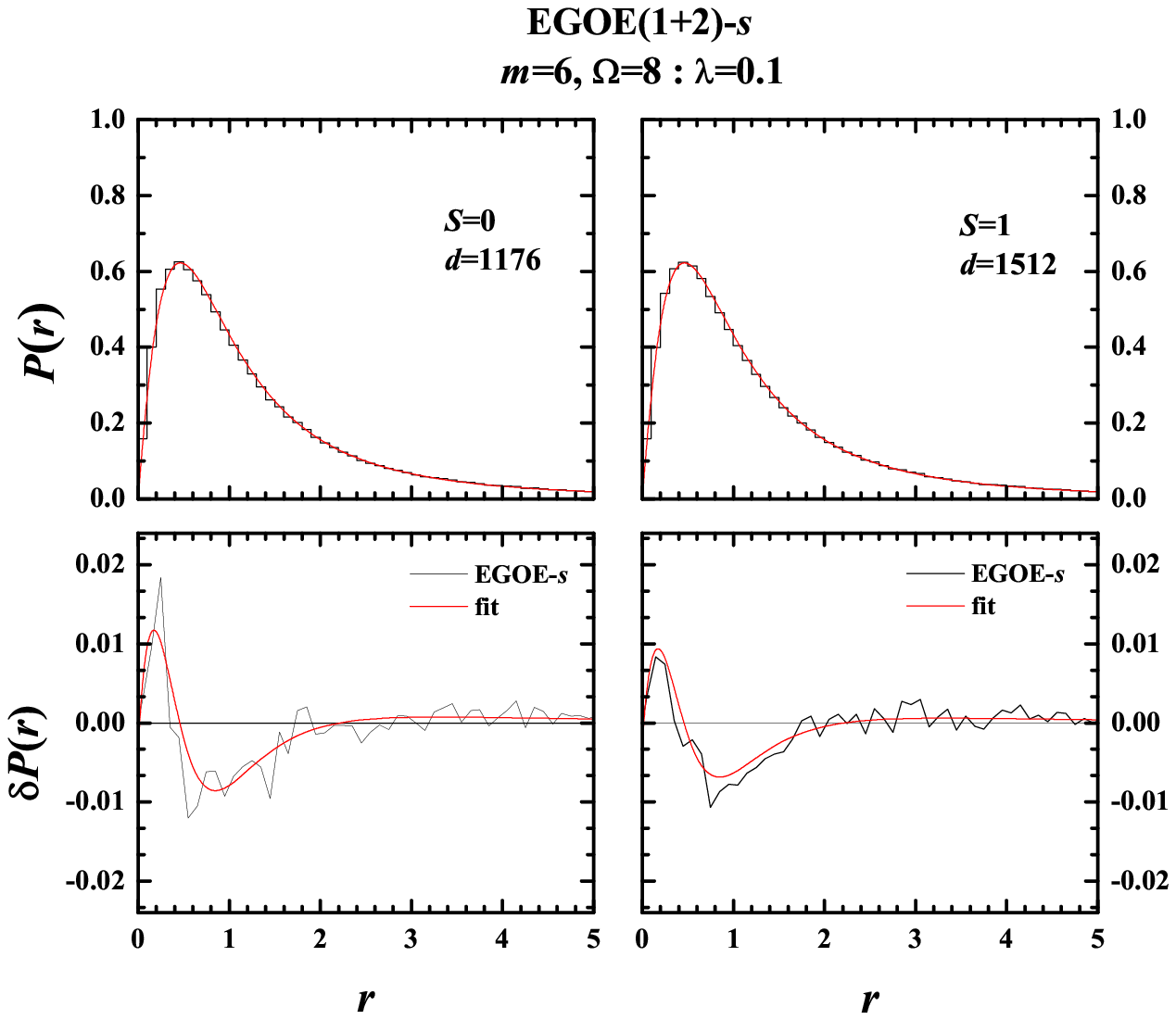}
\end{center}

\caption{Probability distribution of the  ratio of consecutive level spacings
$P(r)$ vs. $r$ and $\delta P(r)$ vs. $r$ for a 500 member EGOE(1+2)-$\cs$
ensemble. Results are shown for spins $S=0$ and $1$. See Fig. \ref{egoe} and
text for details.}

\label{egoe-s}
\end{figure}

\begin{figure}
\begin{center}
\includegraphics[width=0.7\linewidth]{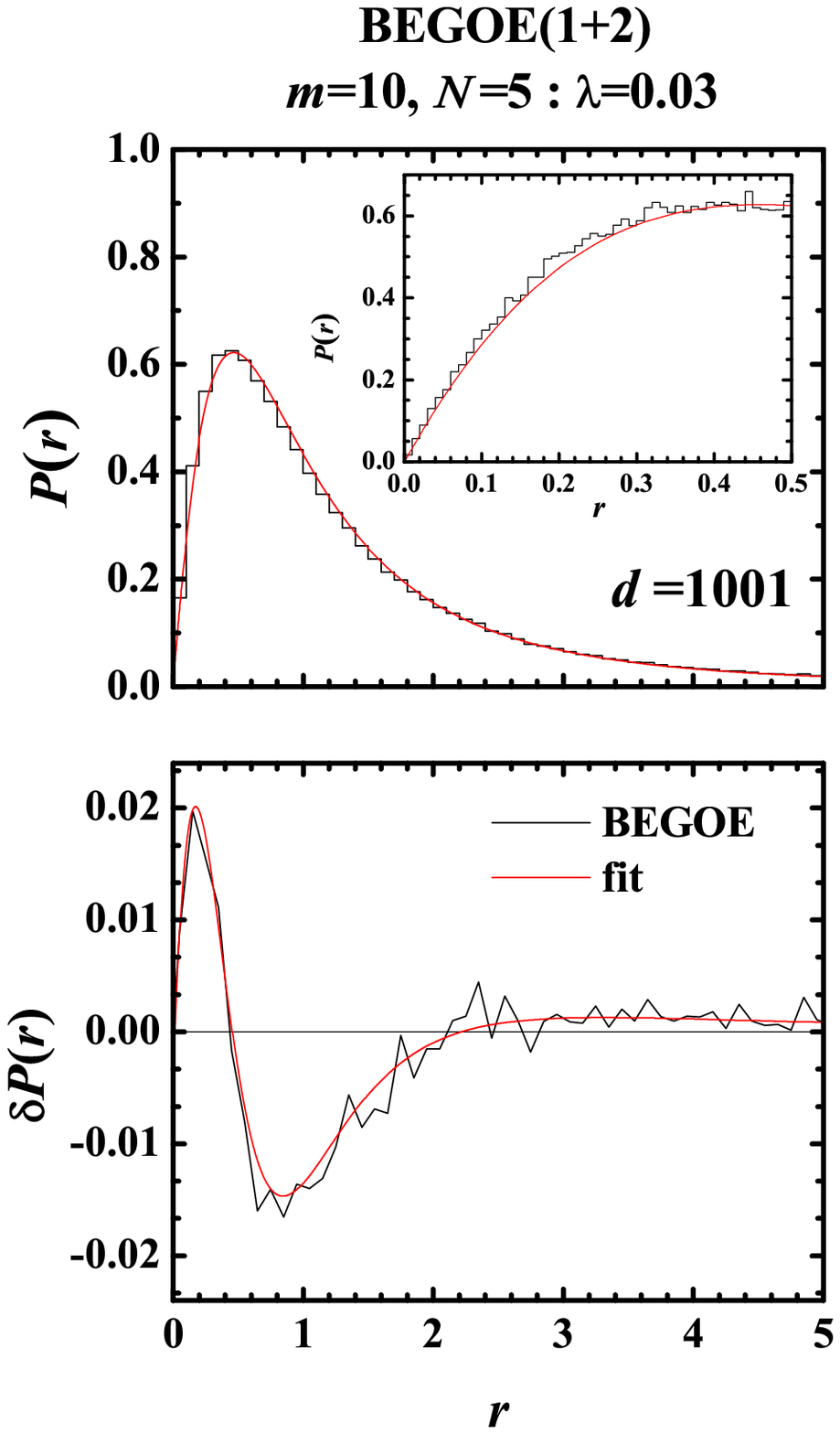}
\end{center}

\caption{Probability distribution of the  ratio of consecutive level spacings
$P(r)$ vs. $r$ and $\delta P(r)$ vs. $r$ for a 500 member BEGOE(1+2) ensemble.
See Fig. \ref{egoe} and text for details.}

\label{begoe}
\end{figure}

\begin{figure}
\begin{center}
\includegraphics[width=\linewidth]{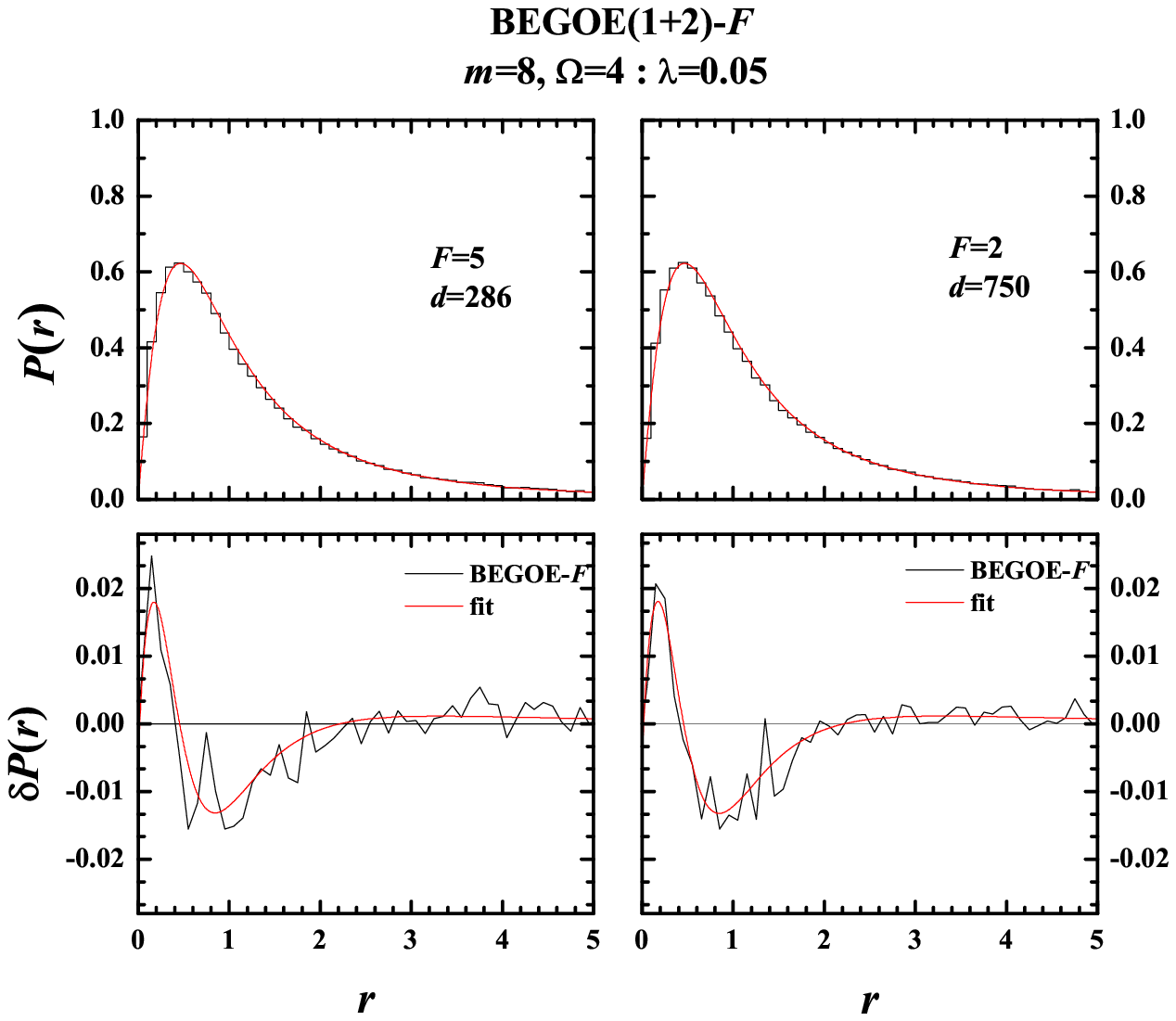}
\end{center}

\caption{Probability distribution of the  ratio of consecutive level spacings
$P(r)$ vs. $r$ and $\delta P(r)$ vs. $r$ for a 500 member BEGOE(1+2)-$F$
ensemble. Results are shown for $F$-spins values $F=5$ and $2$. See Fig.
\ref{egoe} and text for details.} \label{begoe-f}

\end{figure}

\begin{figure}
\begin{center}
\includegraphics[width=0.7\linewidth]{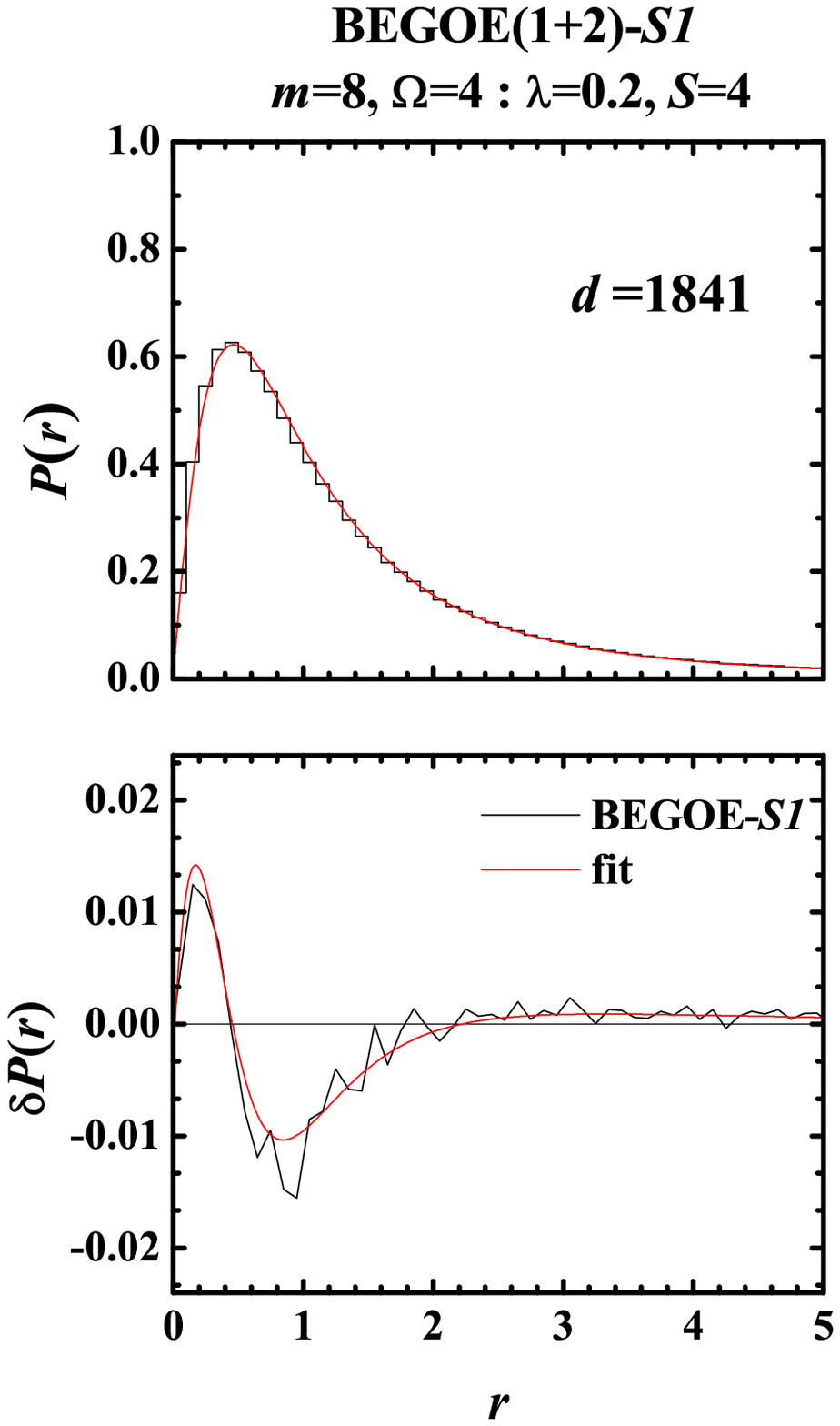}
\end{center}

\caption{Probability distribution of the  ratio of consecutive level spacings
$P(r)$ vs. $r$ and $\delta P(r)$ vs. $r$ for a 500 member BEGOE(1+2)-$S1$
ensemble. See Fig. \ref{egoe} and text for details.}

\label{begoe-s1}
\end{figure}

Following the definitions given in Section \ref{sec:2}, we have constructed the
distribution of the ratio of consecutive level spacings $P(r)$, using the middle
80\% part of the spectrum, for the EEs described in Section \ref{sec:3}. The results are
shown in the Figs. \ref{egoe}--\ref{begoe-s1} (upper panels). In each
calculation, averaging over the 500 members of the ensemble,  histogram for
$P(r)$ is constructed using a bin size of 0.1. The agreement between the
numerical embedded ensemble results and the  surmise given by Eq. (\ref{prw})
is very good for all the examples. Results are  also shown for small value of
$r$ ($r \leq 0.5$) for the EGOE(1+2) and BEGOE(1+2) examples; see the inset
plots in  Figs. \ref{egoe} and \ref{begoe} respectively. They show that the
agreements are indeed very close for $r$ over all the range. We have also
calculated the difference between numerical results and the surmise for all the
examples considered in the study. These results are shown in the lower panels of
the Figs. \ref{egoe}--\ref{begoe-s1}. The smooth red curves (shown in lower
panels) are obtained by fitting Eq. (\ref{diff}) with one parameter $C$. The
values of $C$ are given in Table \ref{table1}. It is seen from Figs. \ref{egoe}--\ref{begoe-s1} that
the $\delta P(r)$ given by Eq. (\ref{diff}) fits quite well the numerical
results. In addition, we have also calculated the averages $\langle r \rangle$
and  $\langle \tilde{r} \rangle$ and the results are given in Table \ref{table1}. Again
the calculated values are seen to be close to GOE values.

\begin{figure}
\begin{center}
\includegraphics[width=0.9\linewidth]{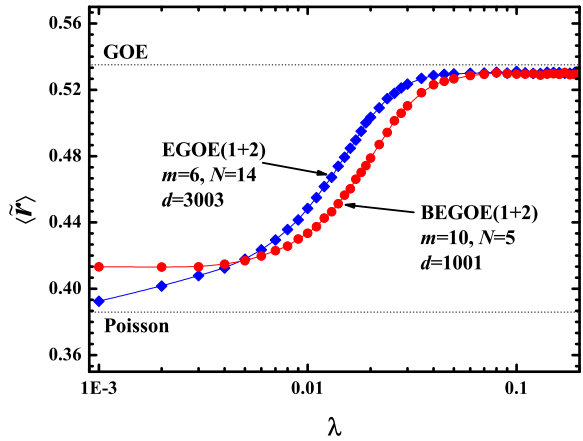}
\end{center}

\caption{Ensemble averaged value of $\tilde{r}_n$ (denoted as $\lan \tilde{r}\ran$) defined by Eq. (\ref{eq.rn}) as a
function of two-body interaction strength $\lambda$, calculated using 500 member
ensembles for EGOE(1+2) ensemble with $(m,N)=(6,14)$ and BEGOE(1+2) ensemble
with $(m,N)=(10,5)$. The horizontal dotted-lines represents Poisson estimate
(bottom reference line) and Wigner estimate (top reference line). In the
calculations sp energies are drawn from the center of a GOE. As the systems
considered are  not strictly Poisson for $\lambda=0$, it is seen from the figure
that for very small $\lambda$ value, the  $\lan \tilde{r}\ran$ value is larger
than the  Poisson value with the difference more for BEGOE(1+2) as $N$ is much
smaller for this example.}

\label{rn}
\end{figure}

In the past it is demonstrated that for EGOE(1+2) and  BEGOE(1+2) ensembles, as
the strength $\lambda$ of the two-body interaction increases, generically there
is Poisson to GOE transition in level fluctuations
\cite{vkbk01,cpk-2003,vyas-12,mkc-2010}. In order to examine this in terms of
$P(r)$,  we have calculated $\langle \tilde{r} \rangle$ for spin-less fermion
EGOE(1+2) and  spin-less boson BEGOE(1+2) ensembles as a function of the
interaction strength $\lambda$. Figure \ref{rn} shows the results. It is clearly
seen that as $\lambda$ increases from $\lambda=0$, the value of $\lan \tilde{r}
\ran$ changes from close to Poisson value to finally the GOE value. Similar
results are obtained using lattice models in \cite{Pal-10,Iyer-12}. In order to
quantify the results in Fig. \ref{rn}, it is necessary to derive a formula for
$P(r)$ in a random matrix model that generates Poisson to GOE transition; see
ahead for further discussion.

\begin{table}
\centering

\caption{Values of the constant $C$ and the averages $\lan \tilde{r} \ran$
and $\lan r \ran $ for various EE examples used  in the present study. Given are
also the values for Poisson, GOE, EGOE(2) and BEGOE(2).}

\begin{tabular}{cccccc}
\hline
\hline
Example & $C$ && $\lan \tilde{r} \ran$ &&$\lan r \ran $\\
\hline
\\
EGOE(1+2) & 0.2198 $\pm$ 0.0109 && 0.5304 && 1.7727\\\\
EGOE(1+2)-$\cs$  &  & & \\
$S=0$ & 0.1909 $\pm$ 0.0119 && 0.5313 && 1.7700\\
$S=1$ & 0.1528 $\pm$ 0.0074 && 0.5318 && 1.7661\\\\
BEGOE(1+2) & 0.3264 $\pm$ 0.0083 && 0.5279 && 1.7960\\\\
BEGOE(1+2)-$F$  &  & & \\
$F=5$ & 0.2926 $\pm$ 0.0159 && 0.5283 && 1.7962\\
$F=2$ & 0.2940 $\pm$ 0.0120 && 0.5286 && 1.7885\\\\
BEGOE(1+2)-$S1$ &  & & \\
$S=4$ & 0.2303 $\pm$ 0.0082 && 0.5303 && 1.7718\\
\\
EGOE(2) & & & &&\\
all levels & 0.2153 $\pm$ 0.0088 && 0.5303 && 1.7823\\\\
\\
BEGOE(2) & & & &&\\
all Levels & 0.2788 $\pm$ 0.0093 && 0.5286 && 1.9340\\\\
\\
GOE & 0.2334 && 0.5359 &&  1.75 \\
\\
Poisson & $\cdots$ && 0.3863 && $\infty$ \\
\\
\hline
\hline
\end{tabular}
\label{table1}
\end{table}

\begin{figure}
\begin{center}
\includegraphics[width=0.7\linewidth]{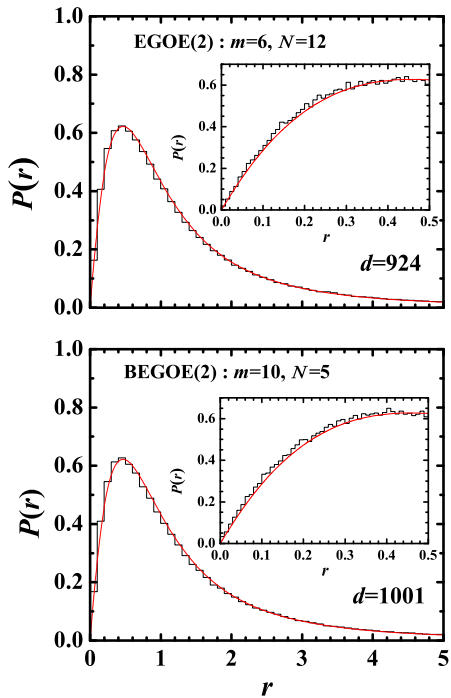}
\end{center}

\caption{Results for $P(r)$ vs $r$ for EGOE(2) and BEGOE(2) examples including
all the levels in the analysis. Inset figures show the results for $r \leq 0.5$.
The red smooth curve is $P_W(r)$ given by Eq. (\ref{prw}).}

\label{full}
\end{figure}

\begin{figure}
\begin{center}
\includegraphics[width=1.1\linewidth]{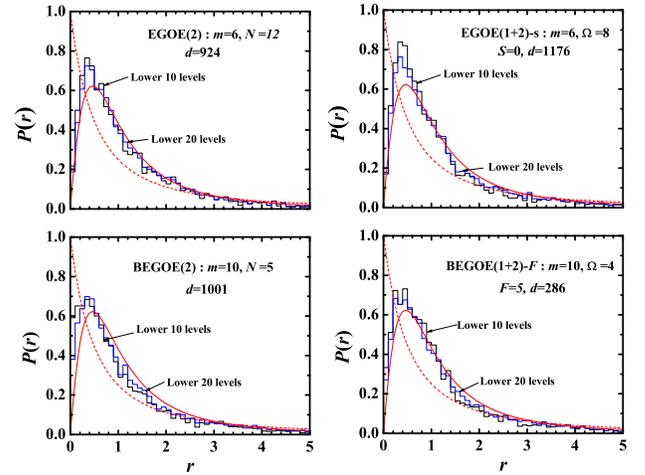}
\end{center}

\caption{Probability distribution of the  ratio of consecutive level spacings
$P(r)$ vs. $r$ for the lowest 10 and 20 levels using for EGOE(2), BEGOE(2),
EGOE(1+2)-$\cs$ and BEGOE(1+2)-$F$ ensembles. For comparison, results from
$P_P(S)$ (red dash curve) and $P_W(S)$ (red smooth curve) are shown.  See text
for details.}
\label{lowlevel}
\end{figure}

\begin{figure}
\begin{center}
\includegraphics[width=\linewidth]{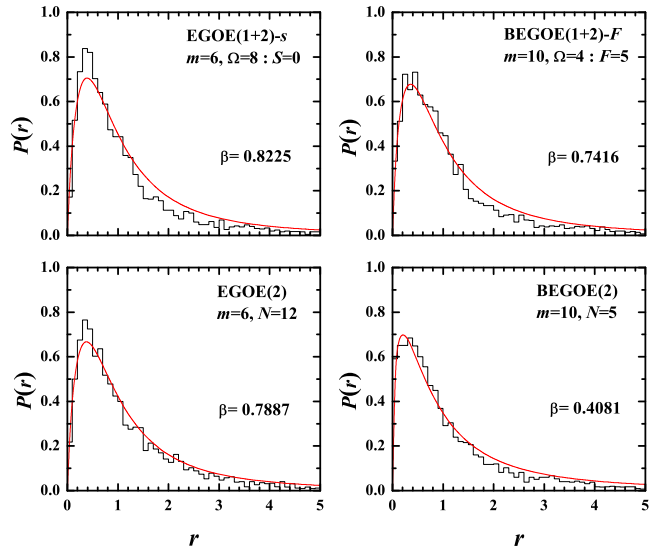}
\end{center}

\caption{Histograms represent $P(r)$ vs. $r$ results for the lowest 10 levels obtained for
EGOE(1+2)-$\cs$, BEGOE(1+2)-$F$  EGOE(2) and  BEGOE(2) examples. The red
smoothed curves are due to fitting histograms with Eq. (\ref{eqpgoe})
and values of parameter $\beta$ are given in the figure. We have also analyzed
the lowest 20 levels for the same examples and the $\beta$ values obtained are :
$0.8399$[EGOE(1+2)-$\cs$], $0.7895$[BEGOE(1+2)-$F$], $0.8795$[EGOE(2)] and $0.6205$[BEGOE(2)].}
\label{fit-low10}
\end{figure}

With all the good agreements seen in Figs. \ref{egoe}-\ref{begoe-s1} for
EGOE(1+2)s with sufficiently large value for the interaction strength and
including middle 80\% of the levels in the analysis, the questions that one may
ask are: (i) will there be deviations if we use all the levels; (ii) do we get
good agreements if we use EE without mean-field; (iii) will the $P_W(r)$ given
by Eq. (\ref{prw}) fit well the results if we use only the lowest 10 or 20 levels.
Briefly, in order to answer (i) and (ii) we have carried out EGOE(2) and
BEGOE(2) calculations using all the levels for $(m=6,N=12)$ and $(m=5,N=10)$
systems respectively. The results in Fig. \ref{full} and Table 1 clearly show
that the agreement with $P_W(r)$ given by Eq. (\ref{prw}) is good even when we
include all the levels in the analysis. Question (iii) above is important as
levels close to the ground state are in general expected to show departures from
GOE. In fact, Flores et al \cite{flores2001} showed, using eigenvalues from
large nuclear shell model calculations, that the so-called semi-Poisson
distribution gives a better fit to the NNSD when only low-lying levels are used
along with spectral unfolding. To investigate deviations from GOE if any present
in low-lying levels generated by EEs,  we have constructed $P(r)$ considering
lowest 10 levels and 20 levels for  EGOE(2), BEGOE(2), EGOE(1+2)-$\cs$  and
BEGOE(1+2)-$F$ examples. Results are shown in Fig. \ref{lowlevel}. There are
clearly deviations from GOE for $r \lazz 0.5$. In order to quantify the
departures from GOE for low-lying levels, it is necessary to extend the results
in \cite{ABGR-2013} to Poisson to GOE interpolating region. Using Eq. (7) of
\cite{ABGR-2013}, we  suggest the following form for Poison
to GOE interpolation for $P(r)$,
\be
  P_{P-GOE}(r:\beta) = \dis\frac{1}{Z_\beta}\;\dis\frac{(r + r^2)^\beta}{
                  \l[1+ (2-\beta)r +r^2\r]^{1+\frac{3}{2}\beta}}\;.
\label{eqpgoe}
\ee
Then, $\beta=0$ gives Poisson and $\beta=1$ GOE correctly. The condition
$\int^{\infty}_0 P(r) dr =1$ gives $Z_\beta$. Eq. (\ref{eqpgoe}) fits well
the results in Fig. \ref{lowlevel} and some examples are shown in Fig. \ref{fit-low10}. This will be
explored in more detail elsewhere.

\begin{figure}
\begin{center}
\includegraphics[width=0.9\linewidth]{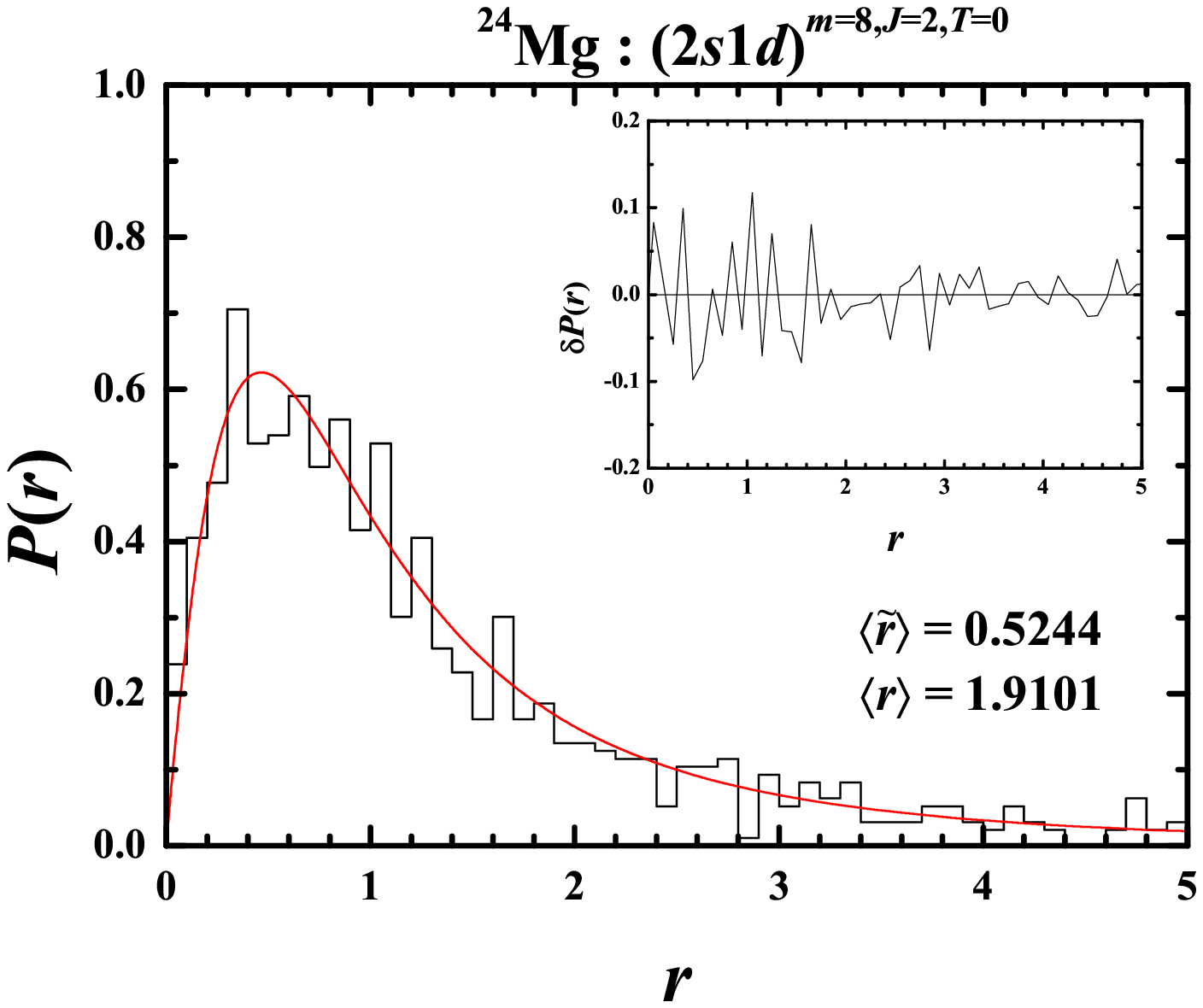}
\end{center}

\caption{Probability distribution of the  ratio of consecutive level spacings
$P(r)$ vs. $r$  for Nuclear shell model example: $^{24}$Mg with 8 nucleons in
the $(2s1d)$ shell with angular momentum $J=2$ and isospin $T=0$. The matrix
dimension is 1206 and all levels are used in the analysis. See Ref.\cite{Lec-08} for further details. Inset figure
shows $\delta P(r)$ vs. $r$. Given in the figure are also the calculated
values of $\lan \tilde{r} \ran$ and  $\lan r\ran$. See Fig. \ref{egoe} and text
for details.}

\label{sm}
\end{figure}

\section{Conclusions}
\label{sec:5}

Embedded ensembles (EGOEs), appropriate for isolated finite quantum
many-particle systems, generate eigenvalue density close to Gaussian and this is
quite different from the GOE semi-circle density. In the past it was shown
\cite{Br-81,Bo-74} that only with proper spectral unfolding,  EGOEs exhibit GOE
level fluctuations. Therefore, nature of level fluctuations in EGOEs is still
not fully understood. Addressing this issue, in this Letter we have
demonstrated, with examples from  EGOE(1+2), EGOE(1+2)-$\cs$, BEGOE(1+2),
BEGOE(1+2)-$F$ and  BEGOE(1+2)-$S1$, that the probability distribution $P(r)$ of
ratio of  consecutive level spacings for embedded random matrix ensembles follow
GOE for strong enough interaction. The Wigner-like surmise for $P(r)$ is found
to agrees very well with the numerical calculations. Also, the difference
between the surmise and the exact calculations is small and can be fitted by a
one-parameter polynomial formula with very good accuracy. As $P(r)$ is
independent of unfolding, the results in Section \ref{sec:4} conclusively
establish that level fluctuations in EGOE(1+2)s follow  GOE for strong enough
interaction. Let us add that we have also verified that $P_W(r)$  is close to
the $P(r)$ from a nuclear shell model example as shown in Fig. \ref{sm}. Note
that the results  from shell model example can be considered as the results of a
typical member  of EGOE(1+2)-$JT$ \cite{Br-81,vkbk01} and this ensemble is
usually called TBRE in literature \cite{PW-07}.  Finally, in future it is useful
to derive a formula for $P(r)$  for Poisson to GOE transition (see \cite{KS} for
a discussion of  this for NNSD) and also for pseudo integrable systems (see
\cite{semipoi} for  NNSD for these systems). These and Eq. (\ref{eqpgoe})
should be useful in quantifying departures from GOE in low-lying levels generated by EEs.

{\center {\bf ACKNOWLEDGMENTS\\}}
We thank V. Potbhare for many useful discussions and for his interest in the
present work. One of the authors (NDC) acknowledges support from UGC (New
Delhi) grant No: F.40-425/2011(SR).

\end{document}